# The Organisation of the Elderly Connectome

Alistair Perry[1,2,3], Wei Wen[1,2], Anton Lord[3], Anbupalam Thalamuthu[1,2], Gloria Roberts[2,4], Philip B. Mitchell[2,4], Perminder S. Sachdev[1,2], Michael Breakspear[2,3,5]

[1]*Centre for Healthy Brain Ageing (CHeBA), School of Psychiatry, University of New South Wales, Sydney, NSW, Australia,*

[2]*School of Psychiatry, University of New South Wales, Sydney, New South Wales 2052, Australia*,

[3]*Systems Neuroscience Group, QIMR Berghofer Medical Research Institute, 300 Herston Road, Herston, QLD 4006, Australia*

[4]*Black Dog Institute, Prince of Wales Hospital, Hospital Road, Randwick, NSW 2031, Australia*

[5]*Metro North Mental Health Service, Royal Brisbane and Women's Hospital, Herston, QLD,* 4029, Australia

*Corresponding author:*

Associate Professor Dr Wei Wen

Euroa Centre, Prince of Wales Hospital, Barker Street, Randwick NSW 2006

Email: w.wen@unsw.edu.au

Phone: +61 2 382 3730

Fax: +61 2 9382 3774

**Abstract**

Investigations of the human connectome have elucidated core features of adult structural networks, particularly the crucial role of hub-regions. However, little is known regarding network organisation of the healthy elderly connectome, a crucial prelude to the systematic study of neurodegenerative disorders. Here, whole-brain probabilistic tractography was performed on high-angular diffusion-weighted images acquired from 114 healthy elderly subjects (age 76-94 years; 64 females). Structural networks were reconstructed between 512 cortical and subcortical brain regions. We sought to investigate the architectural features of hub-regions, as well as left-right asymmetries, and sexual dimorphisms. We observed that the topology of hub-regions is consistent with a young adult population, and previously published adult connectomic data. More importantly, the architectural features of hub connections reflect their ongoing vital role in network communication. We also found substantial sexual dimorphisms, with females exhibiting stronger inter-hemispheric connections between cingulate and prefrontal cortices. Lastly, we demonstrate intriguing left-lateralized subnetworks consistent with the neural circuitry specialised for language and executive functions, while rightward subnetworks were dominant in visual and visuospatial streams. These findings provide insights into healthy brain ageing and provide a benchmark for the study of neurodegenerative disorders such as Alzheimer's disease (AD) and Frontotemporal Dementia (FTD).

## 1. Introduction.

The human brain is a large-scale complex network known as the human "connectome" (Sporns et al., 2005). The application of graph theoretical analysis to human neuroimaging data has uncovered topological features of the connectome that mirror other complex systems (Fornito et al., 2013; Sporns, 2013). These network features include "small-worldness" (Achard et al., 2006; Sporns and Zwi, 2004; Stephan et al., 2000), highly-connected "hubs" (Hagmann et al., 2008; van den Heuvel and Sporns, 2011, 2013b), and a modular structure (Hagmann et al., 2008; Meunier et al., 2009). Knowledge of the connectome has accelerated through recent advances in diffusion-weighted imaging, including optimal acquisition parameters (Sotiropoulos et al., 2013; Tournier et al., 2013), improved reconstruction algorithms (Behrens et al., 2003; Tournier et al., 2010), and diffusion models (Aganj et al., 2011; Behrens et al., 2007; Jbabdi et al., 2012; Tournier et al., 2008).

A crucial architectural feature of the adult human connectome is the presence of highly-connected regions ("hubs"), that are also densely connected with each other (van den Heuvel and Sporns, 2013b). These regions form what is known as a "rich-club", and occur in cortical regions such as the precuneus, cingulum (anterior and posterior), insula, superior frontal and parietal areas, temporal regions, and also subcortical structures (van den Heuvel and Sporns, 2011, 2013b). Rich-club connections in human (Collin et al., 2014; van den Heuvel et al., 2012) , macaque (Harriger et al., 2012) and cat cortices (de Reus and van den Heuvel, 2013) have high topological efficiency, longer anatomical fibers, increased inter-modular connectivity and route a large proportion of network traffic. The structural rich-club may thus act as a central backbone that integrates communication between segregated brain regions (van den Heuvel and Sporns, 2013b). This is exemplified by the disproportionate reduction in network "communicability" and/or "efficiency" when hub-regions or their connections are artificially lesioned (Crossley et al., 2013; de Reus et al., 2014; van den Heuvel and Sporns, 2011) .

These hub-regions overlap with transmodal areas known to be pivotal within-and-between core neurocognitive systems such as the cognitive control, default mode, and salience network (Crossley et al., 2013; Dwyer et al., 2014; Sepulcre et al., 2012; Spreng et al., 2013; Tomasi and Volkow, 2011; Uddin et al., 2011; van den Heuvel and Sporns, 2013a). Interestingly, alterations in functional connectivity of these large-scale systems in elderly populations have been associated with changes in working memory, processing speed and executive functions (Campbell et al., 2012; Damoiseaux et al., 2008; He et al., 2014; Lim et al., 2014; Wang et al., 2010). These disruptions are thus suggestive of topological changes occuring to hub connections with ageing. Hub-regions are also metabolically costly, evident through their increased metabolic expenditure and wiring cost (Collin et al., 2014; Liang et al., 2013; van den Heuvel et al., 2012). This increased energy expenditure of hub-regions further highlights their potential for age-related changes, as their high metabolic cost has been shown to potentially render such regions more vulnerable to pathological processes in neurodegenerative disorders (Crossley et al., 2014; Liang et al., 2013; Tomasi et al., 2013). Indeed, hub-regions have shown to be more likely susceptible to normal ageing processes such as amyloid deposition (Buckner et al., 2009; Toga and Thompson, 2014).

During cognitively demanding tasks, older adults increase their recruitment of contralateral brain regions, suggesting compensatory mechanisms (Cabeza et al., 2002; Davis et al., 2012; Park and Reuter-Lorenz, 2009). Left-hemisphere networks are well known to be dominant in language tasks, whilst the right-hemisphere is associated with visuospatial abilities (Geschwind and Galaburda, 1985; Herve et al., 2013; Toga and Thompson, 2003). Although connectomic investigations (Caeyenberghs and Leemans, 2014; Nielsen et al., 2013; Tomasi and Volkow, 2012b) have examined lateralized organisation at the nodal-level, no study has specifically investigated lateralization of the elderly connectome

Sexual dimorphism has also been an active area of research for the last few decades, with increasing interest from connectomic investigations (Dennis et al., 2013; Duarte-Carvajalino et al., 2012; Gong et al., 2009; Ryman et al., 2014). Across the lifespan, males have been shown to demonstrate greater performance in visuospatial tasks, whilst females excel on verbal tasks (Gur et al., 2012; Hoogendam et al., 2014; Kimura, 2004). Preferential wiring for inter-hemispheric structural connections was recently observed in female adolescents, whilst localised intra-hemispheric connectivity characterises cortical networks in young men (Ingalhalikar et al., 2014). Whether such topological differences persist into late adulthood is not known.

Hitherto, the structural connectomes of healthy elderly populations have been investigated through lifespan longitudinal studies (Betzel et al., 2014; Caeyenberghs and Leemans, 2014; Gong et al., 2009). Whilst these incorporate sufficiently large numbers of subjects across the life span, the number of elderly subjects is invariably modest. The organisation of healthy older connectomes hence remains relatively unknown and has not benefitted from recent advances in the acquisition and analysis of structural connectomes. The present study addresses this gap by characterising network topology in elderly structural connectomes generated from high-angular resolution fiber bundles. For comparative purposes, structural networks of a young adult population (17-30 years old) are also investigated.

## 2. Methods.

### 2.1. Participants.

142 cognitively healthy elderly individuals were drawn from the Sydney Memory and Ageing Study (Tsang et al., 2013). The longitudinal study involves community-dwelling older adults aged 76-94 years, randomly recruited from the electoral roll. Participants in the present study were cognitively healthy, defined as performance on all neuropsychological test measures were within 1.5 standard deviations of normative published mean values (Tsang et al., 2013). Individuals not meeting these criteria, or who were reported to exhibit a decline in daily living activities by an informant, were excluded if they met international consensus criteria for Mild Cognitive Impairment (Winblad et al., 2004), decided by a clinical case panel chaired by neuropsychiatrists, psychogeriatricians, and psychologists (Sachdev et al., 2010). Other exclusion criteria included dementia, mental retardation, schizophrenia, bipolar disorder, multiple sclerosis, motor neuron disease, active malignancy, or inadequate comprehension of English to complete a basic assessment.

### 2.2. Diffusion MRI acquisition.

Diffusion MRI data were acquired from all subjects on a Philips 3T Achieva Quasar Dual MRI scanner (Philips Medical System, Best, The Netherlands), using a single-shot echo-planar imaging (EPI) sequence (TR = 13586 ms, TE = 79 ms). For each diffusion scan, 61 gradient directions (b = 2400 s/mm$^2$) and a non-diffusion-weighted acquisition (b = 0 s/mm$^2$) were acquired over a 96mm$^2$ image matrix (FOV 240 mm x 240 mm$^2$ ); with a slice thickness of 2.5 mm and no gap, yielding 2.5 mm isotropic voxels.

2.3. Diffusion image pre-processing.

The diffusion MRI scan of each participant was visualised within FSLView (Smith et al., 2004). Participants were excluded from the study if their scan revealed the presence of artefact due to motion effects. Twenty-two participants were excluded due to diffusion artefact, along with six others whose networks were not completely connected. We thus analysed the structural connectomes from 50 males and 64 females (see Table 1).

**Table 1**. Demographic Information

| Gender (M/F) | Male (N = 50) | Female (N = 64) |
|---|---|---|
| | Mean ± SD | Mean ± SD |
| Age (years) | 83.35 ± 4.74 | 82.61 ± 4.02 |
| Education (years)* | 13.37 ± 3.87 | 11.57 ± 2.91 |

* $p < .01$ (*t*-test)

To correct for head motion, the gradient direction matrix was rotated using a customised in-house algorithm (Leemans and Jones, 2009; Raffelt et al., 2012). Next, to reduce spatial intensity inhomogenities, intensity normalisation was performed on the b0 image and subsequently applied to all diffusion-weighted (DW) images (Sled et al., 1998). Lastly, a Higher Order Model Outlier Rejection model (Pannek et al., 2012) identified voxels with residual outliers in the DW signal.

2.4. Whole-brain fiber tracking.

We employed the probabilistic streamline algorithm (iFOD2) (Tournier et al., 2010) to generate high-resolution whole-brain fiber tracks until 5 million in number were reached. The orientation of fiber distributions (FOD) were estimated within MRtrix software (Tournier et al., 2012), by performing constrained spherical deconvolution (CSD, $lmax = 8$) of the diffusion signal (Tournier et al., 2008). Using the default parameters for images of such acquisition (step size = 1.25 mm, minimum length = 12.5 mm, max length = 250 mm, FA termination = 0.1, max angle = 45º), iFOD2 tracked the most probable fiber propagations by sampling a probability density function of the FOD at points along each candidate path. iFOD2 has been shown to improve the accuracy of reconstructing high-angular fiber bundles (Tournier et al., 2012) and prevent biases caused by overshoot (Tournier et al., 2010).

## 2.5. Construction of whole-brain structural networks.

The standard AAL template (Tzourio-Mazoyer et al., 2002) was subdivided into 512 cortical and sub-cortical parcellation regions (SI 1 Table 1) of approximately uniform size (Zalesky et al., 2010b). The AAL parcellation is widely used in structural network investigations (Bai et al., 2012; Caeyenberghs and Leemans, 2014; Gong et al., 2009; Lo et al., 2010; Shu et al., 2012; Shu et al., 2011), but does not include information on the WM-GM boundary for each parcel. We note that the echo-planar readout in diffusion acquisition induces geometric distortions within the diffusion image (Holland et al., 2010). Thus, the spatial alignment of anatomical information (i.e. WM-GM boundary) from the T1 image and the diffusion-weighted image are not particularly accurate (Smith et al., 2012), precluding an explicit check of the WM-GM boundary for each parcel. However, following our dense seeding (see below), all parcels in the group connectome had substantial connections. It is hence highly unlikely that any of the parcellated regions do not include a GM-WM boundary, being hence “hidden” from the WM.

Parcellations within subject-space were achieved by employing affine linear registrations within FSL (Smith et al., 2004). First, the parcellated template was co-registered to the MNI T1 2mm brain template. The MNI template was then co-registered to the subject's Fractional Anisotropy (FA) image. The parcellation template (in MNI space) was subsequently transformed into subject-space by applying the transformation matrix generated from registering the MNI template to the FA image.

Within a weighted graph $G_w$, a weighted connection $w_{ij}$ (if $w_{ij} \geq 3$) represents the number of streamlines from region *i* terminating within a 2mm radius of $j$. $w_{ij}$ were adjusted by the mean fiber length between *i* and *j* (Hagmann et al., 2008), as fiber densities are known to be over-estimated in longer fiber bundles (Smith et al., 2013).These weighted networks were rendered sparse by thresholding to preserve the same connection density across subjects. All analyses reported here are on connectomes at 7.5% sparsity, whilst other sparsities (5% and 10%) are reported within supplementary materials. We note that there do not exist reliable benchmarks for human tractography using a parcellation comparable to the present one. Density in anatomical studies from primates and rodents varies greatly according to the anatomical parcelleation and tracing method. The sparsity levels included in the present study are thus guided by prior practise. It is common practice for fMRI, and to a lesser extent structural networks, to implement a variety of threshold levels around 10% (Sporns, 2013). However, by selecting a multiple of thresholds, we ensure the topological distribution of the elderly connectome we report is not biased by the density of the networks (van Wijk et al., 2010). Binary networks were constructed from these sparse weighted networks, by setting all connections to one. Average connectomes of the current population were also generated. Summary of the steps involved in structural brain network reconstruction is illustrated in Fig. 1. All network and surface visualisations were generated using BrainNet Viewer (Xia et al., 2013) and CARET (Van Essen et al., 2001) software packages respectively.

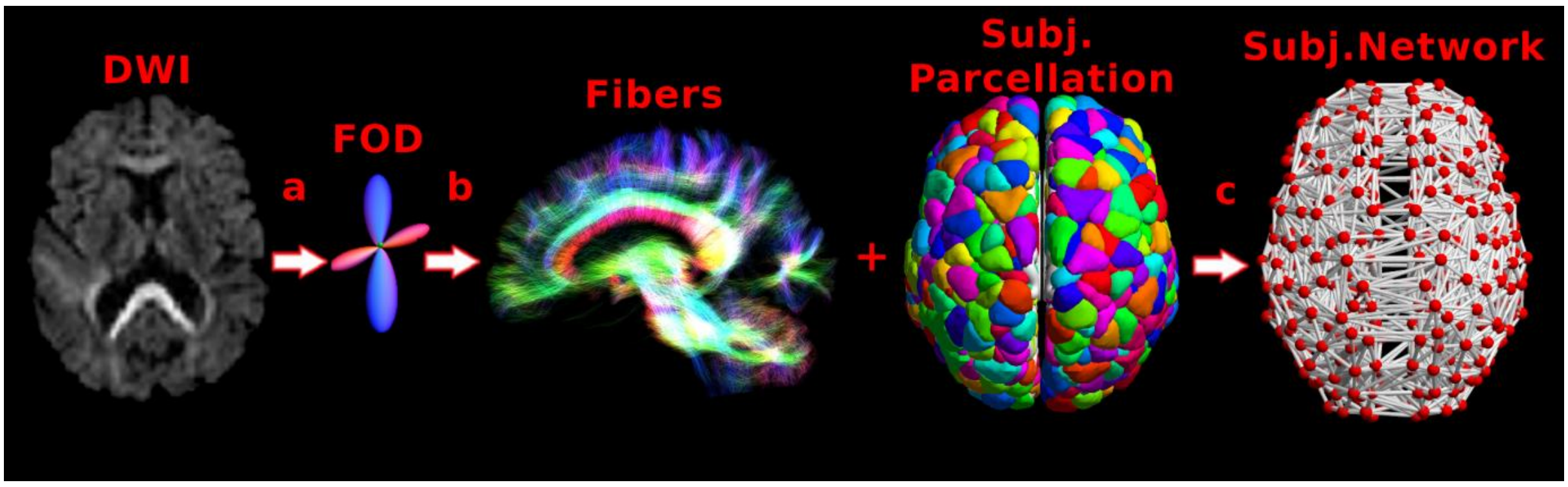

**Fig 1** Steps involved in connectome construction for a representative subject. ***a***, Fiber orientation distributions (FOD) were estimated by performing constrained spherical deconvolution of the diffusion signal within single-fibre voxels of the diffusion-weighted image (DWI). ***b***, High-angular whole-brain fiber tracks were constructed from probabilistic sampling of the FOD. ***c***, Networks representing structural connectivity information generated from the whole-brain fiber tracks were constructed. A connection (white lines) between region *i* and *j* (red dots) of the parcellation (in subject space) was said to be present if a track from *i* terminated within a 2mm radius of *j*.

## 2.6. Graph theoretical characterisation.

### 2.6.1. *Nodal-level measures.*

All nodal-level network measures employed were computed using the Brain Connectivity Toolbox, and have been described elsewhere (Rubinov and Sporns, 2010). Formal definitions are given in SI 2.1.

### 2.6.2. *Community structure*.

A community detection algorithm (Blondel et al., 2008) was employed. The most optimal division of modularity (*Q*) was calculated and a fine-tune tuning algorithm was subsequently employed (Sun et al., 2009). The partition with the highest modularity was retained.

## 2.7. Hub nodes and connection classes.

2.7.1. *Identification of hub-regions*.

Network hubs may be defined according to various network criteria. Here, hub-regions were identified according to aggregate ranking across multiple metrics (Betzel et al., 2014). First, for each subject, each node's "hubness" was calculated from its composite average ranking across degree, betweenness and subgraph centrality scores. The top 15% composite scores (N = 76) were used to identify hub-regions within each subject, whist the top 15% most consistent hubs across subjects were defined as hub-regions across the group.

2.7.2. *Connection classes.* Partitioning nodes into hubs and non-hubs allowed connections to be classified into three types: (1) hub connections, linking hub nodes; (2) feeder connections, linking non-hub nodes to hub nodes; (3) local connections, linking non-hub nodes (van den Heuvel et al., 2012).

2.8. Architectural features of connection classes

2.8.1. *Network density and cost*.

The network cost for each connection was defined as its density (number of streamlines) times its physical length. The network cost for each connection class was calculated as the average cost of its connections. Cost/density ratios for each connection class were calculated as the network cost percentage, divided by its density percentage (van den Heuvel et al., 2012).

2.8.2. *Network Traffic.*

The amount of network traffic along each connection class was based upon the percentage of its connections routing the shortest path between any region *i* and *j*. Here, the shortest path(s) was defined by the minimum number of paths (steps) to reach *j* from *i*, instead of the topological distance (van den Heuvel et al., 2012).

2.9. Network Communicability.

The communicability metric measures the "ease of communication" between *i* and *j*, and is defined by all possible walks of *k* length (steps) between these regions (de Reus et al., 2014; Estrada and Hatano, 2008). Although being a generalisation of shortest path "efficiency" information, the communicability measure does advantageously take into account multiple and longer paths between such regions (de Reus and van den Heuvel, 2014), thus potentially capturing the "parallel processing" nature of brain networks (Alexander and Crutcher, 1990) and thus may be more sensitive to age-related changes impacting upon large number of communication paths. Walks of longer *k* lengths between *i* and *j* have lower contributions to the communicability function than shorter ones, and is defined formally as:

$$C_{ij}= \sum_{k=0}^{k=10}\frac{(G^{k})_{ij}}{k!}=(e^{G})_{ij}\,, \tag{1}$$

where *G* denotes the connectivity matrix, satisfying $G_{ij} = 1$ if region *i* and *j* are connected, and $G_{ij}$= 0 if not. Because a large number of walks can be yielded from large *k* walk lengths between *i* and *j*, $C_{ij}$ was computed until walks of length k = 10. $C_{ij}$ was averaged across all nodal pairs to calculate the overall network communicability within the entire network (de Reus et al., 2014).

## 2.10. Rich-club organisation.

2.10.1. *Rich-club coefficient and significance*. A modified algorithm (Samu et al., 2014) to calculate the weighted richness of hub connections within each subject was implemented. Formal definition of this algorithm, and the examination of the significance of rich-club architecture within these hub connections are detailed in SI 2.2.

## 2.11. Computational attack of hub nodes and their connections

To examine the criticality of hub nodes and their connections to global network communicability, lesions were simulated by randomly removing connections (binary-wise) from each connection class. Because the distribution of connections is by definition unequal across classes (van den Heuvel et al., 2012), the same number of connections were removed across each class for each subject. This action was performed in 25% increments (up until 75%) and average results over 1000 randomly simulated lesions at each increment level were calculated. The change in global network communicability after random edges from each connection class were lesioned, was expressed as a percentage of the intact network's communicability.

## 2.12. Statistical Analysis

2.12.1. *Architectural features of hub-regions and their connections.*

Non-parametric permutation testing was used to assess statistical significance of class differences in connnection metrics (de Reus and van den Heuvel, 2013). First, the difference between the two classes for each subject were computed for a given metric. Second, for each permutation (N = 5000), the metric values were randomly assigned to two random groups and their group difference was computed, resulting in a null distribution of differences. The proportion of the null-distribution values that exceeded the observed original difference was computed and assigned a *p*-value (one-tailed).

2.12.2. *Lateralization and sex differences.*

A general linear model (GLM) was employed to identify differences in weighted edge-wise connectivity. We identified subnetworks that differed significantly between the groups on each effect using the Network-Based Statistic (NBS) software package, which achieves control over type I error (Zalesky et al., 2010a). Networks were permuted 5000 times to obtain the empirical null distribution of the largest network component. A family-wise error (FWE) corrected *p*-value for the network component was estimated by the proportion of permutations for which a network of equal or greater size was identified.

To identify significantly lateralized subnetworks, a repeated measures GLM was employed. For each subject, left intra-hemispheric weighted networks were treated as one condition, whilst right intra-hemispheric networks were treated as the other. For between-group analyses of sexual dimorphisms, age and education level were treated as covariates.

## 3. Results.

### 3.1. Identification of hub-regions and rich-club architecture.

Brain regions identified as hubs, by virtue of having consistently high composite scores across subjects are illustrated in Fig 2. These hubs (for index of surface colours see Fig 2c) are distributed bilaterally in subcortical structures (Fig 2b), and cortical (Fig 2a) regions including the insula, anterior cingulate (AC), precentral gyrus, precuneus, superior frontal, supplementary motor area (SMA), temporal poles, occipital areas, and also the left inferior frontal gryus (IFG). Connections (red lines) between hub-regions are visualised (average connectome) on a circular graphical representation (Irimia et al., 2012; Krzywinski et al., 2009), arranged to their AAL region (Fig 2c). This allows straightforward identification of intra-and-inter-hemispheric connections that exist between hub-regions. The outermost circular bar represents the hub parcellation region (and surface colour), whilst the middle and innermost bars indicate their average composite score (light red to very dark red), and consistency across subjects respectively (light blue to very dark blue). Sub-cortical structures and cortical regions including the AC, insula, left precentral gyrus, left temporal pole, and left IFG have the highest composite scores and greatest consistency across subjects. Weighted rich-club architecture is found to be present ($\Phi[norm]>1$) and significant for these hub connections (Fig 2e) in all subjects (mean $p = < 0.0001$). These hub-regions are consistently top ranked for the nodal metrics used to calculate their composite score (SI Fig 1 and 2), and are also consistently identified as hubs across different sparsity levels (SI Fig 3).

### 3.2. Architectural features of hub nodes and their connections.

Connections with each each subject were classified as either hub (left panel), feeder (middle), or local (right) connections (Fig 3a, visualised here on the average connectome). Local connections accounted predominantly for the cost (left column) and streamline density (right column) across the network, followed by feeders, whilst hub connections comprised only a small percentage (Fig 3b). Non-parametric permutation testing of the cost/density ratios revealed that hub connections are more costly than predicted by their density alone, in comparison to both feeder and local connections *($p < 0.0001$)* (Fig 3b, middle text column). Cost/density ratios of feeders were also significantly greater than the ratio of local connections ($p$ < 0.0001). These patterns of findings were also identified for analyses of mean fiber lengths across connection classes (Fig 3c, $p$ < 0.0001). Feeder connections significantly route the majority of traffic for the shortest communication path (i.e 40% of shortest paths must route through at least one hub) between regions ($p$ < 0.0001, Fig 3d, right column). Hub connections route a significantly greater percentage of traffic than local connections ($p$ < 0.0001). These analyses are consistent at other sparsity levels (SI 1 Fig 4).

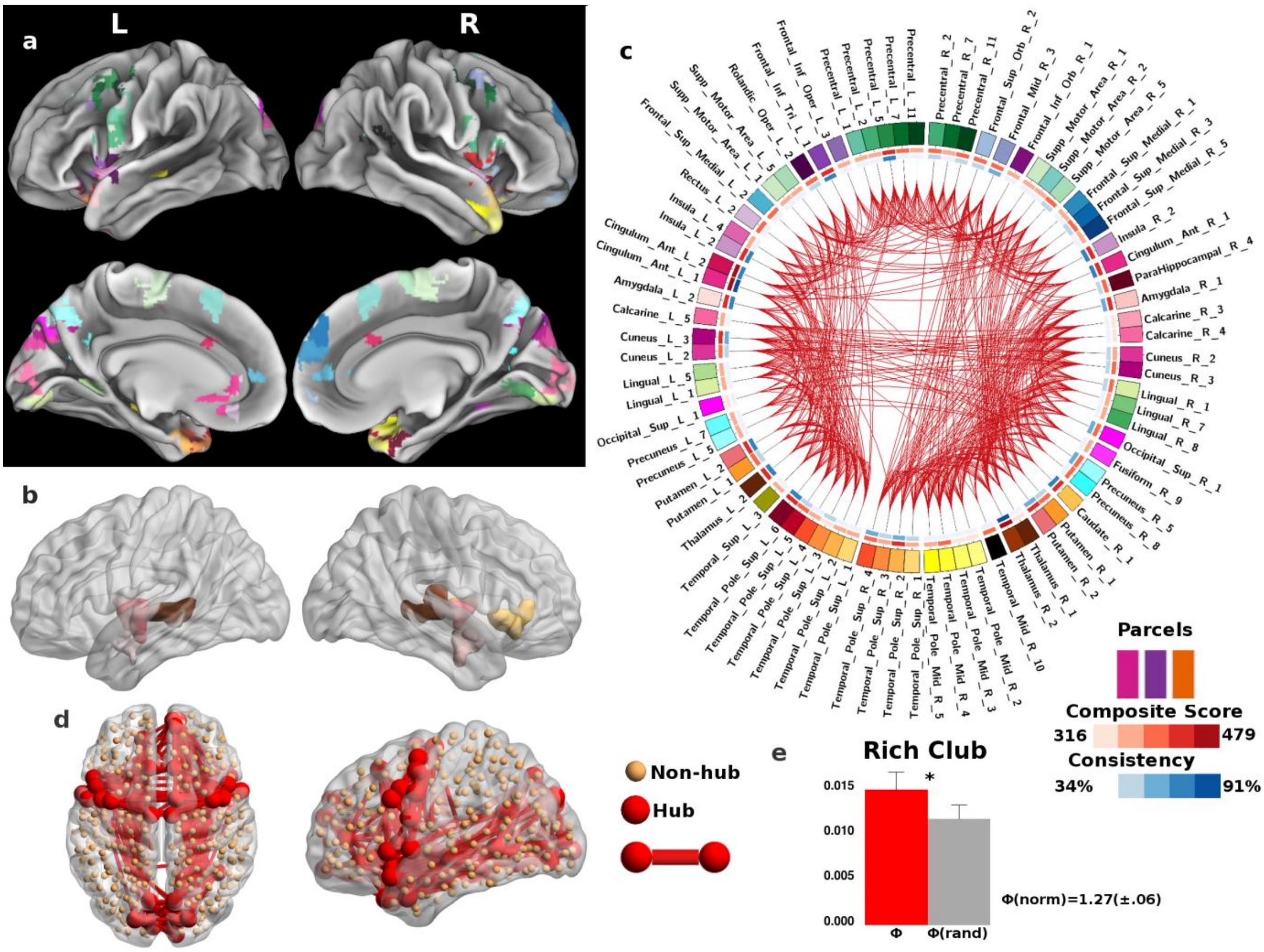

**Fig 2** Brain regions identified to be hubs within the elderly connectome. Surface and volumetric representation of cortical (***a***) and sub-cortical (***b***) hub-regions respectively. Refer to ***c*** for the parcellation regions the surface colours index. **c,** Circular representation of hub-regions and connections among each other (red lines), with regions arranged according to their respective AAL region and hemispheric location (right-hemispheric on the right axis, left on the left). The outermost circular bar represents the hub parcellation region and their index colour (represented in ***a*** and ***b***), whilst the middle and innermost bars indicate their average composite score (light red to very dark red) and consistency across subjects respectively (light blue to very dark blue). ***d,*** Network perspective of connections (red lines) among hub-regions (red dots). ***e***, For these connections, the mean across subjects for the weighted rich-club coefficient Φ(*h*) (red column) , average random rich-club coefficient Φ(*rand*) (grey column), and normalised rich-club coefficient Φ(*norm*). The asterisk denotes weighted rich-club architecture was found to be present (Φ[*norm*]>1) and statistically significant for each subject, as Φ(*h*) were found to be greater than Φ(*rand*) generated from each of the 1000 subject-specific randomised networks.

### 3.3. Computational attack of hub connections.

To examine the role of hub connections to global network, simulated lesioning was performed on each connection class (Fig 3e). The mean percentage change in global network communicability when lesioning hub connections were significantly greater than lesioning the same number of feeder connections ($p$ < 0.0001) except at the 75% increment level. Lesioning local connections had the least impact on communicability ($p$ < 0.0001) at all increment levels.

### **3.4. Comparison to a healthy young adult population.**

To aid the topological comparison of elderly and adult connectomic data, the structural networks of a young adult population were also examined. In order to do so, healthy individuals aged 18-30 years old were drawn from an external study (Roberts et al., 2013). Subjects and acquisition details are provided in SI 3. Whilst acquired on the same MRI scanner as the elderly cohort, there are important differences in the acquisition parameters for this cohort (which are summarised in the Supplementary Methods). To avoid age and gender interactions, will limit the comparison to female subjects only in both cohorts.

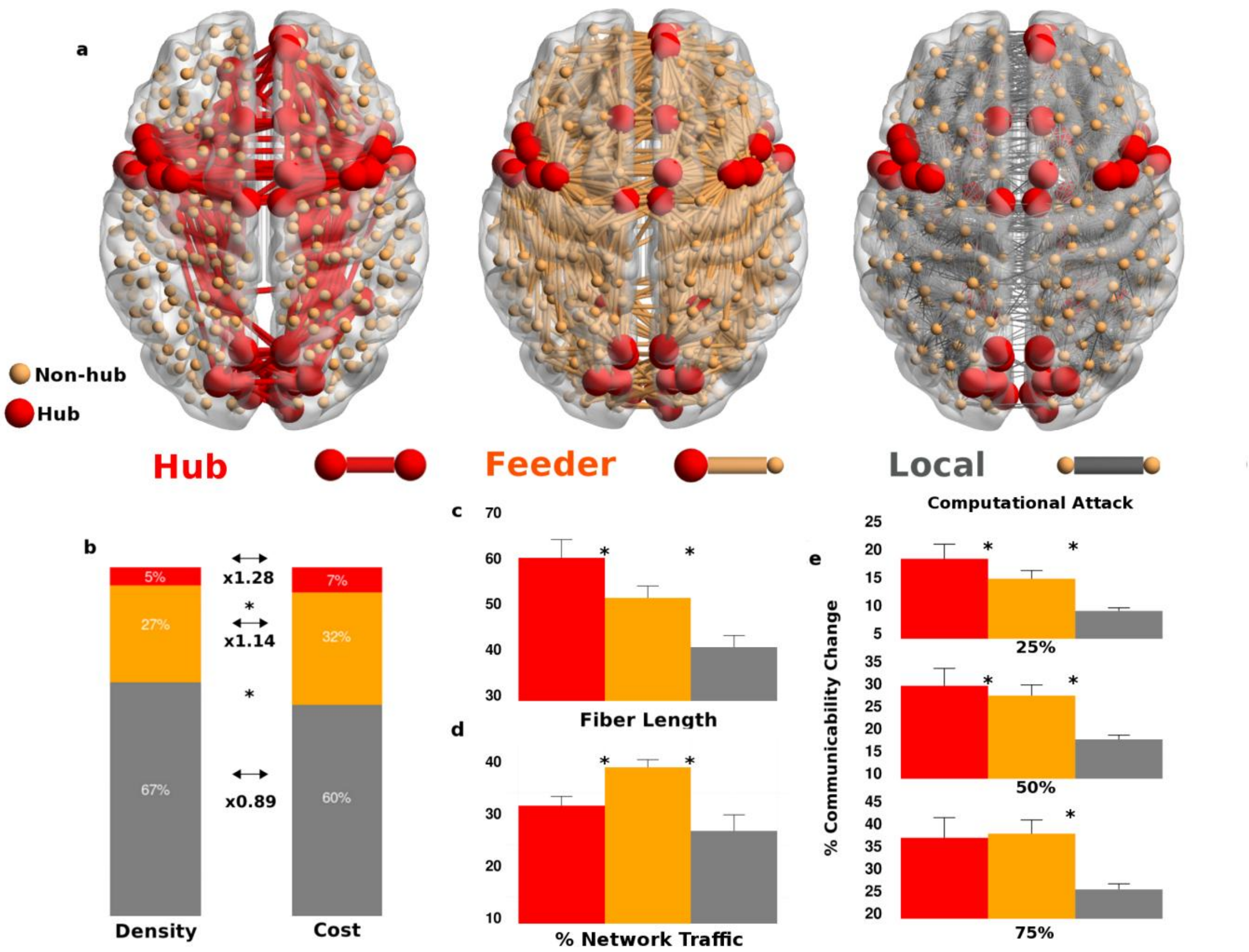

**Fig 3** Architectural features of the different connection classes in the elderly connectome. ***a,*** Connections are classified as either those of hub (left panel) linking (red lines) hub (red dots) regions, feeder connections (middle, orange lines) linking hub to non-hub (orange dots), or local (right, grey lines) connections linking non-hub regions. ***b,*** Mean contributions of each connection class to density (number of streamlines) (left column) and cost to the network (right). The middle text column represents the mean cost/density ratios for each connection class. ***c***, Mean fiber length (mm) for each connection class. ***d***, The mean percentage of network traffic each connection class routes for the shortest path (minimum number of paths) between any region *i* and *j*. ***e,*** Mean percentage change in network communicability after removing specified number of edges from connection class, at 25% (i.e. 25% of rich-club connections) (top), 50% (middle), and 75% (bottom) increments.

* $p < .0001$, permutation testing ($N = 5000$)

The comparison of connection classes and their architectural features across the young and old cohorts is illustrated in Fig 4. Visual inspection of the topological distribution of connection class across the young (Fig 4a) and elderly (Fig 4b) cohorts is dominated by the overall consistent, with only relatively minor qualitative differences evident. In the young cohort, hub connections appear slightly more dispersed, while they also show increased inter-hemispheric connectivity (especially posteriorly) for feeder and local classes (for sagittal perspective, see SI Fig 5b). The architectural features between connection classes in the younger cohort are generally similar to those of the elderly cohort (both male and female), presented in section 3.2 (cf Fig 3, bottom panel). However, several notable differences between the two cohorts are apparent: The mean fiber lengths across the young female cohort (Fig 5d) are markedly longer than the corresponding connection classes in the elderly female cohort (Fig 5g). The proportion of network traffic routed through hub and feeder connections is slightly larger in young females (Fig 5e), whilst traffic routing through local connections is less in the younger relative to the elderly females (Fig 5h).

A direct visual comparison of the distribution of hubs regions in each of these age cohorts is provided in SI Fig. 5a. Those unique to either the young or elderly cohort are given in green and red respectively; Those identified in both populations are given in yellow (SI Fig 5a). The figure again reveals considerable consistency in the distribution hub-regions across the two populations: 72% of hubs identified in elderly females are also hubs in young females. The figure also shows subtle differences between the two cohorts: A cluster of hub nodes unique to the elderly appear within the right temporal pole, left mid-frontal, and prefrontal cortices. Conversely, a cluster of hub regions unique to the younger cohort are found in superior frontal, precentral, and ventral striatal areas.

3.5.Community structure and intra/inter-modular connectivity.

The average elderly connectome (at 7.5% sparsity) partitioned optimally into five distinct modules ($Q$ = 0.67, Fig. 5). The five modules are: left precuneus-occipital-temporal (yellow), left parietal-frontal (green), right frontal-prefrontal (orange), right precuneus-occipital-temporal (red) and a bilateral prefrontal network (blue). Six and five modules were obtained at 5% (SI Fig 6a) and 10% sparsity (SI Fig 6b) respectively. A force-vector algorithm that acts to cluster densely, mutually connected nodes (Jacomy et al., 2014), yields a network perspective of the connectome (Fig. 5c). This reveals the intriguing topological organisation of inter-and-within-module connectivity. The clear division of both hemispheres (division running at an angle), demonstrates the dominance of inter-module intra-hemispheric connectivity. Notably, the integration of the inter-hemispheric community structure is almost entirely achieved through the bilateral prefrontal cortex. In addition, hub-regions (bigger circles) are predominately located along the boundaries of modules, and also embedded within their community affiliation, revealing their intra-and-inter-module connectivity. The mean Pi (inter-module connectivity) and mean Z scores (intra-module connectivity) for hub-nodes were significantly greater than non-hub nodes ($p < 0.0001$, Fig 5d).

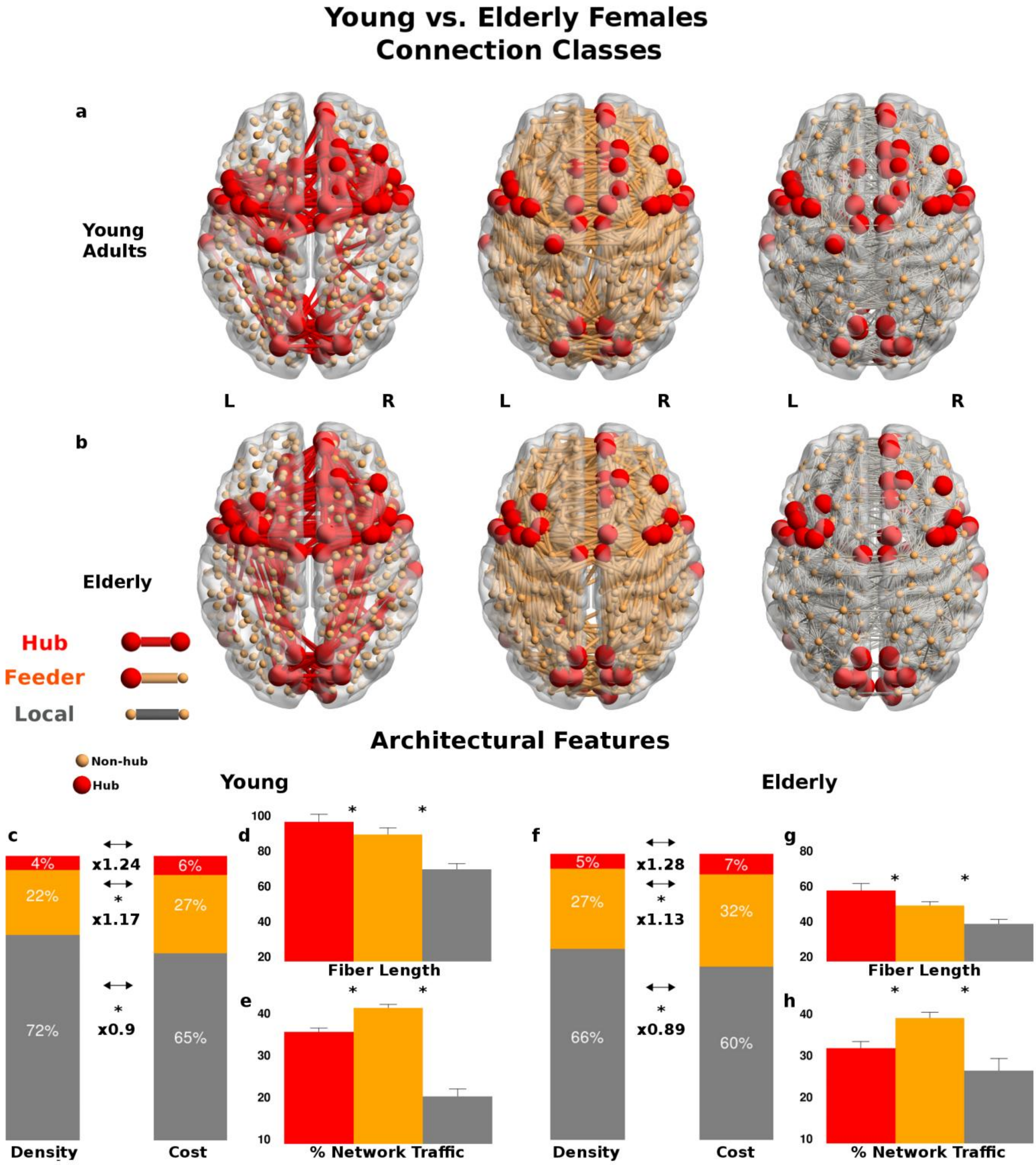

**Fig 4** Comparison of connection classes, and their architectural features, across young adult and elderly females. ***a*** and ***b***, Superior perspective (for sagittal perspective see SI Fig 5a) of connections classified as either those of hub (left panel), feeder (middle), or local (right) connections in the young and elderly connectomes, respectively. ***c*** and ***f***, Mean contributions of each connection class to density (number of streamlines) (left column) and cost to the network (right) in young and elderly females, respectively. The middle text column represents the mean cost/density ratios for each connection class. ***d*** and ***g***, Mean fiber length (mm) for each connection class in young and elderly females, respectively. ***e*** and ***b***, The mean percentage of network traffic each connection class routes for the shortest path (minimum number of paths) between any region *i* and *j*, within young and elderly females, respectively. * $p < .0001$, permutation testing ($N = 5000$)

## 3.6. Lateralization.

Application of the NBS identified significant lateralization of weighted connectivity in two distinct left lateralized clusters ($t = 5.5$, $p < 0.001$, FWE-corrected) and three right-lateralized clusters ($t = 5.5$, $p < 0.0001$, FWE-corrected):

### 3.6.1. *Left-lateralized subnetworks (Fig 6, left panels)*:

Tracking of the fiber bundles corresponding to the first left-lateralized subnetwork reveals a large tract that is consistent with the cingulum and interior fronto-occipital bundles, connecting occipital, precuneus, thalamic, and cingulate structures to orbitofrontal areas (Fig 6a). The second subnetwork involves three distinct components; the first (Fig 6b) consistent with the frontal aslant, connecting the supplementary motor area (SMA) to the inferior frontal operculum; the second (Fig 6c) consistent with the direct arcuate fasiculus, wiring superior temporal and inferior frontal regions; and the third (Fig 6d) connecting temporal (superior and middle) areas to angular and supramarginal regions.

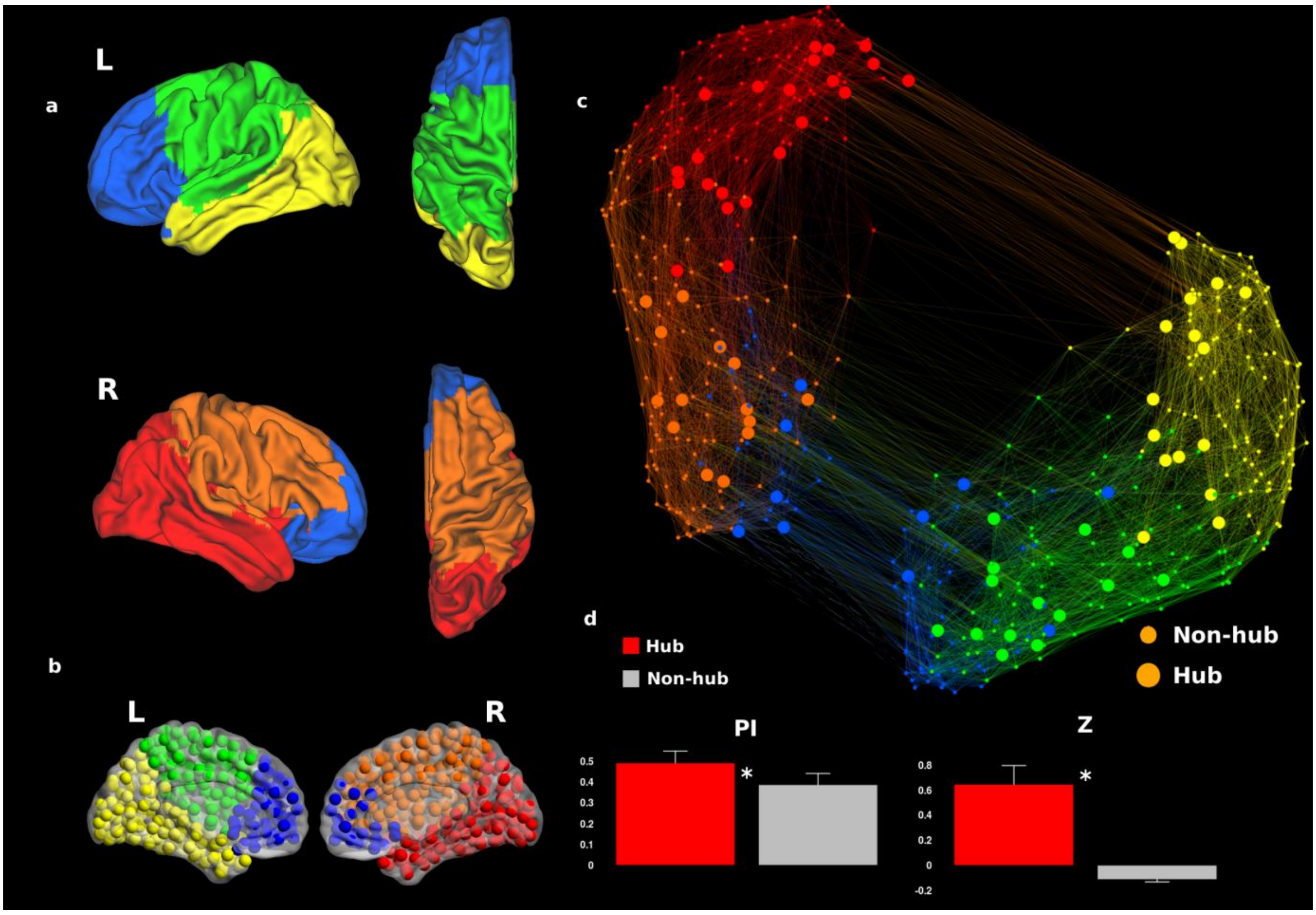

**Fig 5** Community structure and intra-and-inter-module connectivity in the elderly connectome. Surface (***a***), and nodal representation (***b)*** of community structure in the group average connectome, showing the formation of five distinct modules (indexed by different colours). ***c,*** Network perspective of the elderly average connectome through employing a force-vector algorithm, designed to cluster nodes (circles) by virtue of being densely mutually connected. ***d*** shows the organisation of intra-and-inter-module connectivity (lines), with nodal regions of the same community clustered together (intra-module connections coloured according to the community), whilst regions of different communities that are clustered strongly are highly-connected (colours are mixed). ***e,*** Mean PI (inter-module participation, left panel) and Z-scores (intra-module participation, right panel) for hub and non-hub regions.

3.6.2. *Right-lateralized subnetwork (Fig 6, right panels)*:

The first right-lateralised subnetwork invokes two distinct components; one consistent with superior longitudinal circuits spanning from superior temporal regions to the insula and ventral striatum, whilst the second component involving loops between precuneus and occipital regions (Fig 6e). The second subnetwork is consistent with superior longitudinal circuits connecting inferior parietal areas to the insula (Fig 6f). The last subnetwork is consistent with the circuits of optic radiations connecting the thalamus to medial temporal and occipital areas, but also includes connections between temporal (middle and medial) and occipital structures (Fig 6g). All significant subnetworks are listed in SI 4 Table 1 and Table 2 (clusters highlighted in bold).

Each cluster identified above was also significantly lateralized ($p < 0.0001$, permutation testing) within only right-handed subjects ($N = 109$), in concordance with the NBS analyses (SI 1 Fig 7).

3.7. Sexual dimorphism.

We also identified gender-associated subnetworks (Fig 7, SI 4 Table 3; $t = 3.5$, $p < 0.05$, FWE-corrected). Three distinct subnetworks were more strongly expressed in females, all involving increased inter-hemispheric connectivity: The first (Fig 7, top row, $p < 0.015$) includes connection between middle cingulate structures, and also between the left middle cingulate and right SMA. The second (Fig 7a, middle, $p < 0.0001$) includes connections spanning bilateral AC structures, and also AC and superior frontal structures. The third (right, $p < 0.015$) encompasses connections from the left IFG to the right middle and superior frontal regions. The strongest gender-related differences in males (Fig. 7b, red) encompassed two similar subnetwork of connections; The first ($p < 0.004$) encompassed connections within the left-hemisphere, spanning from subcortical (thalamus, putamen) and AC structures to refrontal (orbitofrontal, rectus, and superior medial) cortex. The second ($p < 0.047$), within the right hemisphere, connected the ventral striatum to orbitofrontal cortex, with further connections to frontal superior medial regions.

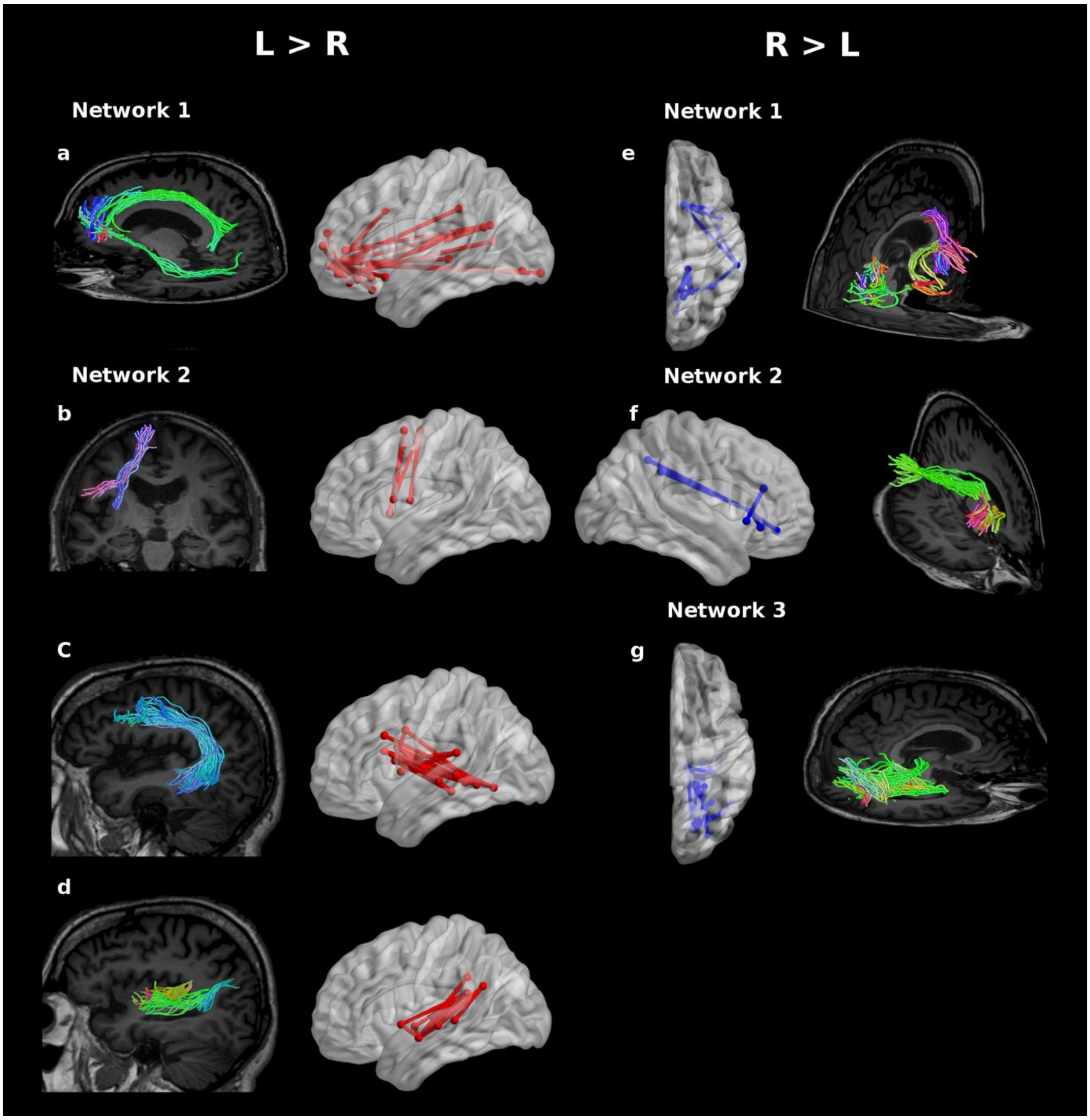

**Fig 6** Subnetwork clusters identified by the NBS ($p < .01$, FWE-corrected). The figure shows two network clusters to be left-lateralized (left panel, ***a, b, c***, and ***d).*** The figure also shows three network clusters to be right-lateralized (right panel, ***e***, ***f***, and ***g***).

## 4. Discussion.

We sought to elucidate key features of the healthy elderly connectome, leveraging recent advances in the acquisition and analysis of brain network connectivity. We found the topology and architectural features of hub-regions to be consistent with connectomic data from a young healthy adult cohort, and also with prior research in young adults (van den Heuvel and Sporns, 2011, 2013b) . We also report the presence of strongly lateralized subnetworks, and focal sexual dimorphisms in network organisation within the elderly connectome.

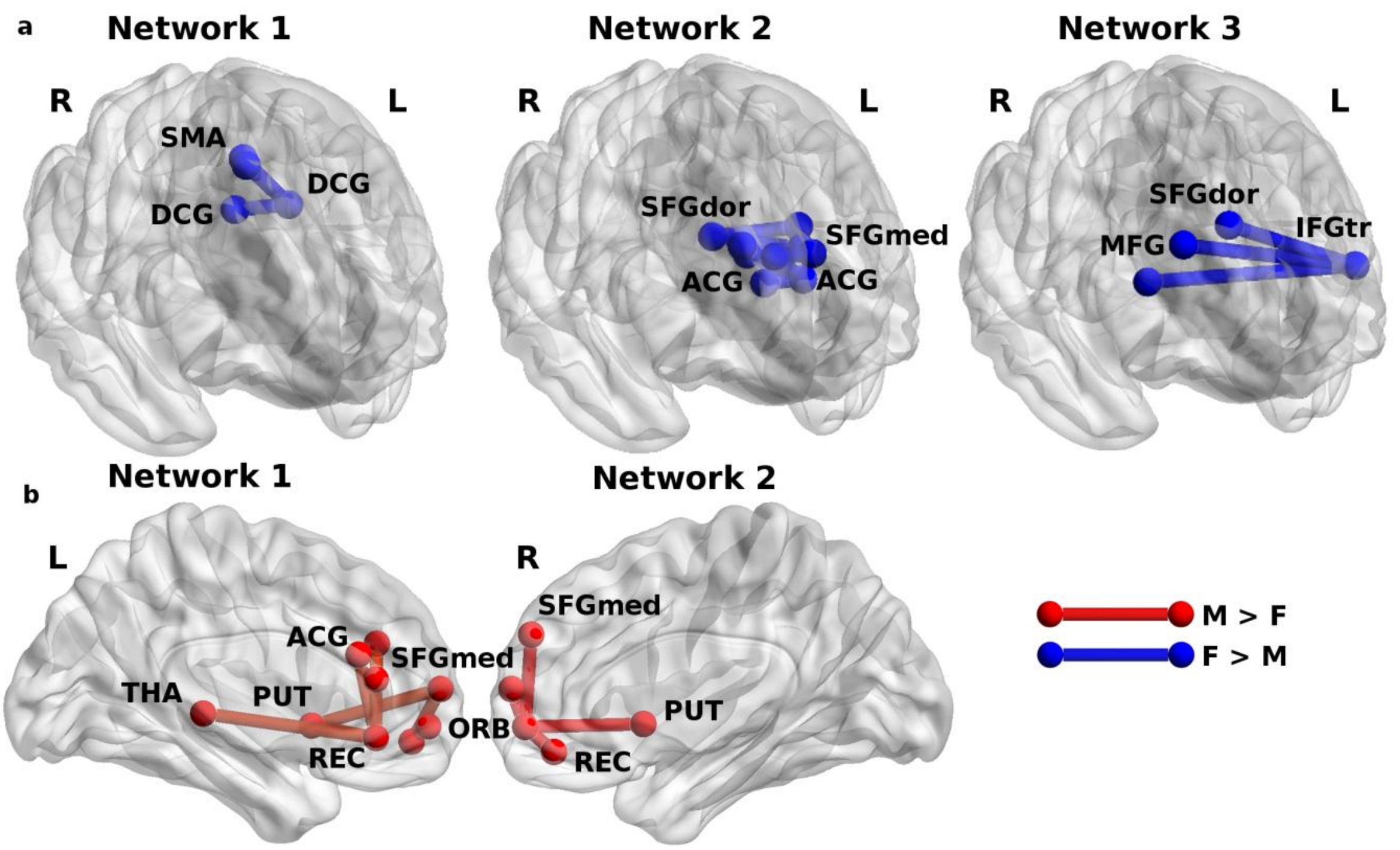

**Fig 7** Focal gender differences in subnetwork connectivity ($p < .05$, FWE-corrected). ***a***, Blue lines represent the localised subnetworks of anatomical connections between nodes (blue dots) where the NBS identified the strongest connectivity greater in females, relative to males. ***b***, Red lines and dots indicates the localised connectivity of subnetworks strongest in males.

SMA, Supplementary Motor Area; DCG; Middle Cingulate, SFGdor, Superior Frontal; SFGmed, medial Superior Frontal; ACG, Anterior Cingulate Gyrus; MFG, Middle Frontal; IFGtri, Inferior Frontal Triangularis ; THA, Thalamus; PUT, Putamen; REC, Rectus; ORB, Orbitofrontal.

## 4.1.Hubs in the elderly connectome.

Hub-regions identified in the elderly connectome are highly consistent with the topology we identified in the structural networks of a young adult population, although subtle differences do occur. Hub-regions identified here have also predominately been revealed as structural hubs - according to various definitions - in other investigations of healthy adults (Betzel et al., 2014; Collin et al., 2014; van den Heuvel et al., 2012; van den Heuvel et al., 2010; van den Heuvel and Sporns, 2011). In our data, superior parietal and posterior cingulate regions were not identified as hubs in either the young or eldery connectomes, in contrast with most previous investigations. Hub-regions in the elderly connectome with the largest composite scores, and also showing the highest consistency across subjects, include subcortical structures (i.e. thalamus, striatum, and the amygdalae), and cortical regions such as the AC, insula, and precentral gyrus. The majority of these regions have been shown to be the most highly connected (both weighted and binary-wise) in other studies of adult structural networks (Collin et al., 2014; van den Heuvel and Sporns, 2011), where their nodal properties rank highly across multiple measures (Betzel et al., 2014; Crossley et al., 2014; van den Heuvel et al., 2010; van den Heuvel and Sporns, 2011). Notably, these are core regions that have been prposed to form the adult "rich club": Densely connected hubs, with enriched inter-hub connectivity suggesting an integral role in large-scale network communication (van den Heuvel and Sporns, 2011, 2013b). Here we also we reveal that the topology of core hub-regions follow a consistent distribution across the healthy lifespan.

Many of the specific architectural features of hub connections are also consistent with young adults (Collin et al., 2014; Crossley et al., 2014; van den Heuvel et al., 2012). Hub connections (including feeders) were found to exhibit longer projection distances and increased cost-to-density ratios (more costly than predicted by their density alone), underlining their likely high-cost to brain networks. Hub connections were also found to exhibit weighted rich-club architecture, route a greater proportion of network traffic (relative to their density), and possess stronger inter-and-intra-modular connectivity. Most notably, virtual lesioning of hub connections were found to result in a disproportionate reduction in global network communicability, in comparison to the removal of feeder (except at the last increment level) and local connections. These high-cost features of hub connections in the elderly appear to be offset by their functional advantages in integrating brain regions of distributed large-scale systems. Given these features are found within our younger cohort, and also previously published adult connectomic data (Collin et al., 2014; van den Heuvel et al., 2012), these findings thus suggests the critical role of hub connections to large-scale network communication is ongoing across the lifespan.

The critical role of hub-regions and their connections to large-scale brain network dynamics is generating wider empirical attention. Hub-regions overlap with multiple large-scale functional networks (Braga et al., 2013; Crossley et al., 2013; Sepulcre et al., 2012; Spreng et al., 2013; Tomasi and Volkow, 2011; van den Heuvel and Sporns, 2013a; Yeo et al., 2014), and their connections have been shown to be involved in a disproportionately greater amount of integration of these networks (van den Heuvel and Sporns, 2013a). Furthermore, hub-regions are predominately those regions important to the integration of dynamic large-scale networks during various cognitive states (Dwyer et al., 2014; Elton and Gao, 2014; Fornito et al., 2012; Sripada et al., 2014), and overlap with areas implicated for their higher-order roles within-and-between such systems (Grahn et al., 2008; Lindquist et al., 2012; Menon and Uddin, 2010; Shackman et al., 2011b; van den Brink et al., 2014). Interestingly, the brain regions typically reported to display these characteristics in large-scale systems are transmodal subcortical (i.e. thalamus, caudate nucleus) and limbic (i.e. insula, anterior cingulate) areas also identified to be most representative of hubs in the elderly connectome (the precuneus was a notable exception in our data). In turn, this provides further plausibility regarding the stability across normal ageing of not only the topology of core architectural brain features, but also their pivotal roles in large-scale network communication.

Despite these similarities across the age cohorts, we did observe some possible age-related changes to hub-regions (and their connections). Notably, the mean fiber length of connections to hub-regions (both hub and feeder connections) in elderly females were found to be 40mm shorter than young females. We also observed an increase in the routing of simulated traffic in the elderly connectome through local connections and a corresponding decrease of hub-to-hub routing. These findings can be interpreted within the hallmarks of normal ageing. First, previous investigations of functional connectomic lifespan changes reportedthat long-distance connections are disproportionally affected in normal ageing (Cao et al., 2014; Tomasi and Volkow, 2012a; Wang et al., 2012). Second, cognitive domains (i.e. working memory, executive functions, processing speed) that consistently decline with healthy ageing rely on the integration and coordination of distributed large-scale systems, where long-distance connections are pivotal (Crossley et al., 2013; Dwyer et al., 2014; Lim et al., 2014; Park and Reuter-Lorenz, 2009; van den Heuvel and Sporns, 2013a). Finally, the fragility of healthy adult brain networks to simulated (computational) attack of hub connections has been posited to reflect pathogenic processes (i.e. amyloid deposition) in normal ageing and underlying neurodegenerative disorders such as AD, where the high metabolic activity of such regions has shown to render them more susceptible (Buckner et al., 2009; Crossley et al., 2014; Toga and Thompson, 2014; Tomasi et al., 2013). The present findings, taken together with the literature regarding normal ageing, suggest while the core architectural features of hub connections remain pivotal in the elderly, their capacity for large-scale communication is reduced.

Several methodological challenges do limit the implications for direct age-related analysis. First, for direct statistical contrasts to be performed, the diffusion acquisition parameters should be identical between the two populations; otherwise the distribution of connectivity, regardless of age effects, will be non-uniform (Tournier et al., 2013; Vaessen et al., 2010; Zalesky et al., 2010b). Changes in the b-value, for example will impact upon the diffusion signal to noise ratio – changes that will likely propagate through the diffusion pipeline leading to systematic differences (such as the distribution of inferred fibre lengths). Data from the young adult population employed here were chosen on the merits that the diffusion images were acquired on exactly the same MRI scanner. We do not, however, perform direct contrasts, but limit our comparison to a quantitative visualization. Second, differences in head motion, brain volume, white matter volume, brain anatomy and challenges in the appropriate matching of education and general medical issues are other substantial challenges that require substantial future work before direct comparisons between young and elderly connectomes can be confidentally made.

4.2. Lateralization effects.

The first left-lateralized subnetwork cluster we identified in the elderly is consistent with the cingulum bundle and inferior fronto-occipital fibers, connecting occipital, precuneus, thalamic and cingulate structures to orbitofrontal regions. This is consistent with the left-lateralized FA values commonly found within anterior portions of the adult cingulum bundle (Takao et al., 2013). This lateralization is notable given segments of the cingulum bundle, and orbitofrontal structures, are thought to be essential for executive functions including decision-making and emotional processing (Grabenhorst and Rolls, 2011; Heilbronner and Haber, 2014; Schoenbaum et al., 2009; Shackman et al., 2011a).

The first bundle of the second left-lateralized subnetwork interconnects parietal (angular and Geschwind's area) regions with temporal (middle and superior) areas, and the second is consistent with arcuate fasciculus circuits connecting Wernicke's to Broca's area. These are perisylvian circuits specialised for language (Catani et al., 2005; Price, 2012). The other bundle is consistent with the frontal aslant connecting the SMA to the inferior frontal operculum, which has been reported to be leftward lateralized in adults (Catani et al., 2012; Vergani et al., 2014). The SMA and rolandic areas are activated during the movements essential for speech production (Bouchard et al., 2013; Brown et al., 2009; Price, 2012), thereby suggesting this subnetwork is specialised for sensorimotor integration. Nevertheless, these strong leftward lateralizations are surprising given decreased functional specialisation of both prefrontal and language networks is typically reported with age (Antonenko et al., 2013; Bergerbest et al., 2009; Cabeza et al., 2002; Davis et al., 2012).

We also identified three rightward lateralizations. The third of these is associated with visual circuits, consistent with the optic radiation wiring the thalamus to both occipital and medial temporal regions (Bassi et al., 2008; Bürgel et al., 1999; Thiebaut de Schotten et al., 2011b). The two other right-lateralized subnetworks are both consistent with superior longitudinal fasiculus bundles spanning from supramarginal and superior temporal regions to the insula and ventral striatum. These bundles are found to be right-lateralized in adults (Thiebaut de Schotten et al., 2011a; Thiebaut de Schotten et al., 2011b), but of more significance is that the degree of lateralization in these circuits has recently been associated with increased speed for visuospatial processing for targets in the left hemifield (Thiebaut de Schotten et al., 2011a).

These findings suggest these lateralized subnetworks remain specialised for visual and visuospatial processes in the elderly.

4.3. Sexual dimorphism.

We not only replicate findings of increased inter-hemispheric connectivity within female youths (Ingalhalikar et al., 2014), but show it is localised to subnetworks of circuits wiring cingulate structures (middle and anterior), as well as prefrontal cortices (lateral and middle). This observation builds upon prior evidence of distinct sexual dismorphisms within these anatomical areas, such as increased grey matter volume in the prefrontal cortices of females (Feis et al., 2013; Luders and Toga, 2010), as well as greater FA values within the corpus callosum (CC) (Kanaan et al., 2012; Phillips et al., 2013; Schmithorst et al., 2008). Nevertheless, focal identification of these subnetworks is of interest, given that language and executive functions are associated with the same circuits (Gasquoine, 2013; Koechlin et al., 1999; Price, 2012), and females across all age groups demonstrate greater performance in cognitive tasks assessing these functions (Gur et al., 1999; Hoogendam et al., 2014; Kimura, 2004). Furthermore, increased FA of the CC has been associated with increased behavioural performance and inter-hemispheric functional connectivity during language-based tasks (Antonenko et al., 2013; Davis et al., 2012). Thus, it is possible the increased connectivity of these subnetworks in females facilitates superior performance in verbal-based abilities.

We also find stronger connectivity in males in subnetworks connecting ventral striatal, AC and prefrontal regions (orbitofrontal and superior medial). The subnetworks encompass circuits that have been attributed to decision-making and regulatory functions (Basten et al., 2010; Grabenhorst and Rolls, 2011; Winecoff et al., 2013; Zald and Andreotti, 2010).

Interestingly, males generally demonstrate more efficient behavioural regulation, and also differential functional activation in these areas for tasks involving emotion processing and decision-making (Lighthall et al., 2012; Ross and Monnot, 2011; van den Bos et al., 2013; Whittle et al., 2011). This pattern of stronger wiring found in males is thus consistent with the observed gender differences in tasks associated with these circuits.

## 5. Conclusion.

In sum, our study is the first systematic investigation of network organisation in the elderly connectome. Notwithstanding the methodological caveats highlighted above, we provide preliminary evidence that the topology and architectural features of hub-regions are preserved into the healthy elderly. Moreover, our findings provide a benchmark for future longitudinal and clinical investigations, arguing that elucidating the topology and cost of hub regions may be key to connectomic changes. In particular the architectural features shown here provide a benchmark for further connectomic investigations to dissociate healthy ageing from neurodegenerative disorders.

## 6. Acknowledgements.

This work was supported by National Health and Medical Research Council Program (350833). Address correspondence to Wei Wen, Euroa Centre, Prince of Wales Hospital, Barker Street, Randwick NSW 2031. Email: w.wen@unsw.edu.au

## 7. References.